\documentclass[letterpaper, 10 pt, conference]{ieeeconf} 
\IEEEoverridecommandlockouts
\usepackage{amsmath,amssymb,amsfonts}
\usepackage{algorithmic}
\usepackage{booktabs}
\usepackage{multirow}
\usepackage{hhline}
\usepackage{cite}
\usepackage[utf8]{inputenc}
\usepackage[T1]{fontenc}
\usepackage{graphicx}
\usepackage{textcomp}
\usepackage{comment}
\usepackage{amsthm}
\usepackage{xcolor}
\usepackage{breqn}
\usepackage{color}
\usepackage{lipsum,cuted}
\usepackage[ruled,vlined]{algorithm2e}
\usepackage{dsfont}
\usepackage{soul}
\newtheorem{theorem}{Theorem}
\newtheorem{problem}{Problem}
\newtheorem{remark}{Remark}

\newtheorem{assumption}{Assumption}

\newtheorem{lemma}{Lemma}

\newcommand{\bA}{{\mathbf A}}
\newcommand{\bB}{{\mathbf B}}

\newcommand{\bX}{{\mathbf X}}

\newcommand{\bx}{{\mathbf x}}
\newcommand{\by}{{\mathbf y}}
\newcommand{\bg}{{\mathbf g}}

\newcommand{\bu}{{\mathbf u}}

\newcommand{\bPsi}{{\mathbf \Psi}}
\newcommand{\bs}{{\mathbf s}}
\newcommand{\bff}{{\mathbf f}}

\usepackage{float}
\usepackage{amsmath}

\begin{document}

\title{\bf Synthesizing Controller for Safe Navigation using Control Density Function}
\author{Joseph Moyalan, Sriram S.K.S Narayanan, Andrew Zheng,  and Umesh Vaidya
\thanks{Financial support from of NSF CPS award 1932458 and NSF
2031573 is greatly acknowledged. Joseph Moyalan, Sriram S.K.S Narayanan, Andrew Zheng, and Umesh Vaidya are with the Department of Mechanical Engineering, Clemson University, Clemson, SC; {\{jmoyala,sriramk,azheng,uvaidya\}@clemson.edu}.}
}

\maketitle

\begin{abstract}
We consider the problem of navigating a nonlinear dynamical system from some initial set to some target set while avoiding collision with an unsafe set. We extend the concept of density function to control density function (CDF) for solving navigation problems with safety constraints. The occupancy-based interpretation of the measure associated with the density function is instrumental in imposing the safety constraints. The navigation problem with safety constraints is formulated as a quadratic program (QP) using CDF. The existing approach using the control barrier function (CBF) also formulates the navigation problem with safety constraints as QP.
One of the main advantages of the proposed QP using CDF compared to QP formulated using CBF is that both the convergence/stability and safety can be combined and imposed using the CDF. Simulation results involving the Duffing oscillator and safe navigation of Dubin car models are provided to verify the main findings of the paper.

\end{abstract}

\section{Introduction}
Most control system applications in robotics and automotive engineering include driving the nonlinear system dynamics from an initial set to a target set while avoiding certain unsafe sets. Safe navigation is a well-known problem in the robotics community, with applications extending to aerospace, unmanned ground vehicles, manufacturing, power systems, etc. Existing literature on the safe navigation problem involves jointly solving the safety and stability problem using artificial potential methods \cite{khatib1986real}. However, using attractive and repulsive potentials often leads to local minima, which is a well-known problem \cite{krogh1984generalized,rimon1990exact}. Another approach includes \cite{romdlony2014uniting}, which makes use of the control Lyapunov function (CLF) and control barrier function (CBF) to design a feedback law to achieve simultaneously both convergence to a target set and avoidance of an unsafe set. However, the control design is problem-specific and less intuitive, and if the safety and convergence objectives conflict, then such a feedback law can't be designed. Another approach uses the combination of the CLF and CBF to construct a quadratic program (QP) to solve for the desired safe control \cite{ames2016control,ames2019control}. However, finding CLF and CBF to formulate the QP is not trivial.\\
The solution to the navigation problem is also achieved in the dual-density space. The density function was first introduced in \cite{rantzer2001dual} as a dual to Lyapunov function for stability analysis. This density function was later used as a safety certificate using the sum of squares optimization method for the analysis and safety of the nonlinear system \cite{rantzer2004analysis}. In \cite{vaidya2018optimal}, the authors utilized navigation measures for designing safe controllers using a convex formulation. In \cite{chen2020optimal}, the authors formulate the safety problem as a co-design problem of finding the density function and optimal safe controller. However, the iterative approach of estimating the density function from the optimal control can be computationally expensive, depending upon the complexity of the problem. In contrast, we design a controller for a given known density function, which is constructed based on the occupancy-based interpretation of the density. Similarly, the use of linear operators such as Koopman and Perron-Frobenius (P-F) operators for convex data-driven approaches for optimal control and control with safety constraints have been explored in \cite{ma2019optimal,huang2022convex,yu2022data,moyalan2023data}. The convex data-driven approaches heavily rely on observable functions to lift the dynamics to the space of functions. The number of observables required to lift the dynamics to function space increases with an increase in the dimension or complexity of the underlying nonlinear system, which makes it computationally expensive. As such, these approaches suffer the curse of dimensionality. In \cite{zheng2023safe}, to avoid the curse of dimensionality, the authors provide an analytical expression for the density-based feedback controller for the navigation of single integrator dynamics, which can be viewed as a dual construction of the navigation functions from \cite{rimon1990exact}. \\
The control design in \cite{zheng2023safe} is only applicable to single integrator dynamics or systems, which can be reformulated as a single integrator using a change of coordinates and inverse dynamics approach. This paper extends the density-based controller introduced in \cite{zheng2023safe} to nonlinear systems with drift. Therefore, unlike \cite{zheng2023safe}, our approach can be applied to any general nonlinear system that contains drift. In this paper, we introduce the notion of control density function (CDF). The CDF is an extension of the density function, just like CLF is the extension of the Lyapunov function for control systems. The construction of CDF is provided based on \cite{zheng2023safe}, which is then utilized to formulate the problem statement as a QP with CDF-based constraints. The CDF-based constraint ensures a nonlinear control system's simultaneous convergence and safety. This contrasts with the approaches given in \cite{ames2016control,ames2019control}  where one needs to augment CLF constraints with CBF to ensure convergence. Another difference includes almost everywhere convergence of the system dynamics, which is a weaker notion of convergence than control Lyapunov functions. As a result, the control law obtained from CDF constraints ensures convergence for almost all initial conditions. The set of all initial conditions not converging to the target set will have a zero Lebesgue measure \cite{moyalan2021sum}. We also provide an example of an underactuated system in the form of the Dubin car model, where obstacles are only present in the subspace of the system dynamics. Finally, unlike \cite{zheng2023safe}, we also show that the density-based safe controller can be combined with a nominal controller for optimal performance by modifying the cost function to include the norm of the difference between the desired safe control and the nominal control. The rest of the paper is organized as follows. Section II contains the problem statement and some preliminaries, and Section III consists of the paper's main results. Section IV discusses the computational framework for the prescribed QP. In Section V, we provide some simulation results followed by some conclusions.

\section{Preliminaries and Problem Statement}
Consider the dynamical system of the form
\begin{align}
    \dot{\bx} = \bff(\bx) + \bg(\bx)\bu\label{eq:dyn_sys}
\end{align}
where $\bx \in D \subseteq \mathds{R}^n$ and $\bu \in U \subseteq \mathds{R}^m$ are the states and the control inputs respectively. We also assume that  $\bff,\bg \in \mathcal{C}^1(D,\mathds{R}^n)$ are continuously differentiable functions on $D$. The unsafe sets are represented by $\mathcal{U} \subset D$. Next, $\bX_0,\bX_T \subset D$ represents the initial set and target set, respectively. In this paper, we assume the target is located at the origin, i.e., $\bX_T = {0}$. We represent $s_t(\bx)$ to be the solution of \eqref{eq:dyn_sys} with respect to some control $\bu$ at time $t$ starting from the initial condition $\bx$. We also denote $\mathcal{B}(D)$ to be the Borel $\sigma$-algebra on $D$ and ${\cal M}(D)$ as the vector space of real-valued measures on ${\cal B}(D)$ and $m(\cdot)$ denotes Lebesgue measure. Also, we represent $\bar{D} := D\setminus \mathcal{N}_{\eta}$ where $\mathcal{N}_{\eta}$ represents a small neighborhood of $\eta$ radius around $\bX_T$. The notation $\nabla_{\bx}$ denotes $[\frac{\partial}{\partial x_1},\dots,\frac{\partial}{\partial x_n}]^\top$ where $\bx \in \mathds{R}^n$. Also, $tr(A)$ represents the trace of the matrix $A$.

\begin{problem}\label{problem1}(Almost everywhere (a.e.) safe navigation)
The primary objective of this paper is to design a control $\bu$ to drive the trajectories of the system given by \eqref{eq:dyn_sys} from almost all initial conditions (w.r.t. Lebesgue measure) from the initial set $\bX_0$ to a target set $\bX_T$ while avoiding the unsafe set $\mathcal{U}$.
\end{problem}

\subsection{Density function for safe navigation}\label{sec:safe_density}
The construction of the density-based safe control for the dynamics given in \eqref{eq:dyn_sys} is inspired by \cite{zheng2023safe}. We define the unsafe set $\mathcal{U}$. Let there be $K$ number of obstacles. We define a continuous scalar-valued function $c_k(\bx)$ for $k=1,\dots,K$. Now, each obstacle set can be defined as follows:
\begin{align}
    C_k := \{\bx\in D: c_k(\bx) \le 0\} \label{eq:obs_set}
\end{align}
Therefore, the set defining the total unsafe region is given by $\mathcal{U} := \bigcup_{k=1}^K \;C_k$. Similarly, we will define another continuous scalar-valued function $b_k(\bx)$ to define the sensing region for each obstacle set as follows:
\begin{align}
    B_k := \{\bx \in D:b_k(\bx) \le 0 \} \setminus C_k
    \label{eq:sensing_set}
\end{align}
Now, we use the functions $c_k(\bx)$ and $b_k(\bx)$ to formulate a smooth inverse bump function. First, we start by constructing the following functions,
\begin{align*}
    m_k &= \frac{c_k(\bx)}{c_k(\bx)-b_k(\bx)},\\
    \psi_k(\bx) &= \frac{\exp(\frac{-1}{m_k})}{\exp(\frac{-1}{m_k})+\exp(\frac{-1}{1-m_k})}.
\end{align*}
Using the functions $m_k(\bx)$ and $\psi_k(\bx)$, we define a piece-wise smooth inverse bump function $\Psi_k(\bx)$ as follows:
\begin{align}\label{eq:bump}
    &\Psi_k(\bx) = \begin{cases}
        0, &\bx \in C_k \\
        \psi_k(\bx), & \bx \in B_k \\
        1, & \text{otherwise}
    \end{cases}.
\end{align}
Note that $\prod_{k=1}^K\bPsi_k(\bx)$ encodes the unsafe set $\mathcal{U}$. To encode information about the target set $\bX_T$, we use 
$V(\bx)$ which acts as a distance function from the current state $\bx$ to the target state $\bX_T$. The $V(\bx)$ can be modified to adjust to the geometry of the underlying configuration space. In the Euclidean space with $\bx \in \mathds{R}^n$, we use $V(\bx) = (\bx-\bX_T)^\top P (\bx-\bX_T)$ for some $P>0$.  Finally, the density function $\rho(\bx)$ for the safe control is given as follows:
\begin{eqnarray}
    \rho(\bx) = \frac{\prod_{k=1}^K\bPsi_k(\bx)}{V(\bx)^{\alpha}} \label{eq:rho_S}
\end{eqnarray}
for some $\alpha >0$.

\begin{figure}[ht]
  \centering
  \includegraphics[width=1\linewidth]{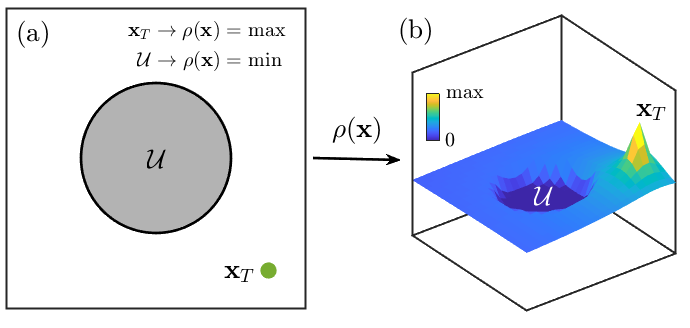}
\caption{(a) Environment setup with unsafe set $\mathcal{U}$ and target $\bX_T$, (b) density function $\rho(\bx)$ for this environment.}
\label{fig:density_function}
\end{figure}

It can be observed that the density function $\rho(\bx)$ given in \eqref{eq:rho_S} is a smooth function for all $\bx \in D$. Fig. \ref{fig:density_function}a shows an environment with one obstacle set $\mathcal{U}$ and target $\bX_T$. Fig. \ref{fig:density_function}b shows the corresponding density function representation. Note that the density function $\rho(\bx)$ takes minimum value for $\bx \in \mathcal{U}$ and max value for $\bx \in \bX_T$.

\begin{assumption}\label{assume:2}
We assume that system dynamics given by \eqref{eq:dyn_sys} is locally linearizable inside the region $\mathcal{N}_{\eta}$ containing $\bX_T=0$. We also assume that the linearized dynamics at the origin are stabilizable. Therefore the local stability of $\bX_T=0$ can be achieved by designing a local linear feedback controller. 
\end{assumption}

The construction of density function in  \eqref{eq:rho_S} was proposed in \cite{zheng2023safe} and used for solving the safe navigation problem with simple integrator dynamics of the form $\dot \bx=\bu$. The feedback controller $\bu$ for the safe navigation was shown to be of the form 
\begin{align}
    \bu(\bx) = \frac{\partial \rho}{\partial \bx} \label{eq:grad_rho}
\end{align}
This paper's main contribution is introducing the CDF for solving the navigation problem as stated in Problem \ref{problem1} for nonlinear system \eqref{eq:dyn_sys} with drift.



\section{Control Density Function (CDF)}
The density functions can be utilized to check the stability of any nonlinear system \cite{rajaram2010stability,moyalan2023data}. However, such density functions cannot be used to design a safe control. By drawing inspiration from how Lyapunov functions were extended to CLF \cite{artstein1983stabilization,sontag1983lyapunov}, and how barrier functions were extended to CBF \cite{wieland2007constructive,ames2019control}, we propose in this section an extension of the density function and call it as the CDF. Let us again consider the nonlinear system as follows:
\begin{align}
    \dot{\bx} = \bff(\bx) + \bg(\bx)\bu\label{eq:ctrl_sys}
\end{align}
where $\bu = [u_1,\dots,u_m]^\top$ and $\bg(\bx) = [g_1(\bx),\dots,g_m(\bx)]$. All the vector fields are assumed to belong to $\mathcal{C}^1(\bar{D},\mathds{R}^n)$.

The following theorem provides the main result for the safe navigation control for a nonlinear system using CDF.
\begin{theorem}\label{thm:1}
    Under Assumption \ref{assume:2}, given system dynamics in \eqref{eq:dyn_sys} and density function given by \eqref{eq:rho_S}, the system trajectories can be driven from almost all initial conditions to a target set $\bX_T$ while avoiding unsafe set $\mathcal{U}$ if there exists a control $\bu \in U$ and $\lambda > 0$ such that
    \begin{align}
   \;\;\;& \nabla \cdot (\bff (\bx)\rho + \bg(\bx)\bu\rho) \ge 0,\;\;a.e.\;\bx\in\Bar{D}\nonumber\\ 
    & \nabla \cdot (\bff (\bx)\rho + \bg(\bx)\bu\rho) \ge \lambda > 0,\;\;\forall\;\bx\in\bX_0\label{eq:theorem_1}
    \end{align}
    The proof is provided in the Appendix.
\end{theorem}

Based on the result of Theorem~\ref{thm:1}, we see that solving for the desired safe control $\bu$ is an infinite-dimensional feasibility problem.
However, if we assume that $\bff(\bx)$, $\bg(\bx)$ and $\rho(\bx)$ are known to us, then we can solve for $\bu$ point-wise along the system trajectory by formulating it as QP utilizing the CDF constraints given in \eqref{eq:theorem_1}. The QP-CDF formulation is given below:
\begin{align}
        &\;\;\;\;\;\;\;\;\min_{\bu}\; \|\bu\|^2 \nonumber \\
       \text{s.t.}\;\;\;& \nabla \cdot (\bff (\bx)\rho + \bg(\bx)\bu\rho) \ge 0,\;\;a.e.\;\bx\in\Bar{D}\nonumber\\ 
    & \nabla \cdot (\bff (\bx)\rho + \bg(\bx)\bu\rho) \ge \lambda > 0,\;\;\forall\;\bx\in\bX_0\label{eq:QP-CDF}
\end{align}
\begin{remark}
    If there exists a nominal control, $\bu_0$, for the system given by \eqref{eq:ctrl_sys}, we can reformulate \eqref{eq:QP-CDF} to enforce $\bu_0$ in the absence of unsafe sets by introducing a cost function which minimizes $\|\bu - \bu_0\|^2$.
\label{Remark1}
\end{remark}

 \section{Computational framework}
The inequality in \eqref{eq:theorem_1} can be expanded as follows:
\begin{eqnarray}
    \nabla \cdot (\bff (\bx)\rho) + \nabla_{\bx}^\top \left[\bg(\bx)\rho\right] \bu + tr\left((\nabla_{\bx}\bu^\top)^\top \bg\rho\right) >0 \label{eq:split_constarint}
\end{eqnarray}
We observe that while trying to solve for $\bu$ point-wise along the system trajectory, the term $\nabla u_i$ is a spatial operator and requires information on points in the neighborhood of the system trajectory. Therefore, to calculate $\nabla u_i$, we perturb the point on the trajectory by $\epsilon$ along the directional basis to obtain points $[z_1,\dots,z_n]$ where $z_i \in \mathds{R}^n$. Here, $z_i = \bx + \epsilon\;\textbf{e}_i$ where $\textbf{e}_i$ is a column vector consisting of all zeros except at $i^{th}$ position where the value is $1$. These points around the trajectory will be used to calculate the $\nabla u_i$.

To make things easier, we can split the inequality given in \eqref{eq:split_constarint} as follows:
\begin{align}
    \nabla \cdot (\bff\rho) + \nabla_{\bx}^\top \left[\bg(\bx)\rho\right] \bu \ge \beta \rho \nonumber\\
    \left|tr\left((\nabla_{\bx}\bu^\top)^\top \bg\right)\right| < \beta   \label{eq:inequality_split}
\end{align}
Now we can write $\nabla u_i$ as follows:
\begin{align}
    \nabla u_i = \left[\frac{u_i^1-u_i}{\epsilon},\dots,\frac{u_i^n-u_i}{\epsilon}\right]^\top \label{eq:grad_u}
\end{align}
where $\bu^j = [u^j_1,\dots,u^j_m]^\top$ represents the control satisfying \eqref{eq:inequality_split} at point $z_j$. This can be written in matrix form for all the $m$ control values as follows:
\begin{align}
    (\nabla_{\bx}\bu^\top)^\top = 
    \begin{bmatrix}
        \frac{u_1^1-u_1}{\epsilon}&\dots&\frac{u_1^n-u_1}{\epsilon}\\
        &\ddots&\\
        \frac{u_m^1-u_m}{\epsilon}&\dots&\frac{u_m^n-u_m}{\epsilon}
    \end{bmatrix} \label{eq:grad_u_matrix}
\end{align}
Therefore, we can rewrite \eqref{eq:inequality_split} as follows:
\begin{align}
   &\nabla \cdot (\bff(\bx)\rho(\bx)) + \nabla_{\bx}^\top \left[\bg(\bx)\rho(\bx) \right]\bu \ge \beta \rho(\bx) \nonumber\\ 
   &\nabla \cdot (\bff(z_1)\rho(z_1)) + \nabla_{\bx}^\top \left[\bg(z_1)\rho(z_1)\right] \bu^1 \ge \beta \rho(z_1) \nonumber\\
    &  \;\;\;\;\;\;\;\;\;\;\;\;\;\;\;\;\;\;\;\;\;\;\;\;\;\;\;\;\;\;\;\;\;\;\;\;\vdots \nonumber\\
   &\nabla \cdot (\bff(z_n)\rho(z_n)) + \nabla_{\bx}^\top \left[\bg(z_n)\rho(z_n)\right] \bu^n \ge \beta \rho(z_n) \nonumber\\
    &\left|tr((\nabla_{\bx}\bu^\top)^\top \;\bg(\bx))\right| < \beta
\end{align}
Here, the number of decision variables to solve the above linear inequalities will be $m(n+1)$.
The algorithm \ref{algo_1} summarizes the steps to solve for safe navigation control using QP-CDF as given in \eqref{eq:QP-CDF}.

\begin{remark}
    The value of $\alpha$ determines the rate of convergence towards the target set. This stems from the fact that $\alpha$ appears in the distance function $V(\bx)$ which contains information on the target set. It can be inferred that the convergence rate can be increased by increasing the value of $\alpha$ and vice-versa. Similarly, increasing the sensing region around the obstacle will lead to smoother avoidance control, and decreasing the sensing region will lead to more aggressive avoidance control near the obstacle. 
\end{remark}


\begin{algorithm}\label{algo_1}
\caption{QP-CDF}
\textbf{Input:} $\bff,\bg,\rho,\bx_0,\beta,N$\\
\For{$k=1:N$}{
$z_j = \bx_{k-1} + \textbf{e}_j\;\;\forall\;j=1,\dots,n$.\\
    \textbf{Solve} \For{$\bu_k,\bu^1_k,\dots,\bu^n_k$}{
    $\min \;\| \bu_k\|^2+\|\bu^1_k\|^2 + \dots+\| \bu^n_k\|^2$\\
    s.t. \\
    $\nabla \cdot (\bff(\bx_{k-1}) \rho(\bx_{k-1}))+$\\
    \quad \quad \quad \quad $\nabla_{\bx}^\top \left[\bg(\bx_{k-1})\rho(\bx_{k-1})\right] \bu_k>\beta \rho(\bx_{k-1})$,\\
    \vspace{2mm}
    $\nabla \cdot (\bff(z_1) \rho(z_1)+\nabla_{\bx}^\top \left[\bg(z_1)\rho(z_1)\right] \bu^1_k>\beta \rho(z_1)$,\\
    \vspace{2mm}
    $\;\;\;\;\;\;\;\vdots$ \\
    \vspace{2mm}
    $\nabla \cdot (\bff(z_n) \rho(z_n)+\nabla_{\bx}^\top \left[\bg(z_n)\rho(z_n)\right] \bu^n_k>\beta \rho(z_n)$,\\
    \vspace{2mm}
     $\left|tr((\nabla_{\bx}\bu_k^\top)^\top \;\bg(\bx_{k-1}))\right| < \beta$ 
    }
$\bx_k = \bx_{k-1} + \Delta t (\bff(\bx_{k-1}) + \bg(\bx_{k-1})\bu_k)$
}
\end{algorithm}

\section{Simulation Results}
In this section, we provide some navigation results for different types of system dynamics starting with the Duffing oscillator.

 \subsection{Duffing oscillator}
 Let us consider the Duffing oscillator dynamics:
 \begin{align}
    &\dot{x}_1 = x_2 \nonumber\\
    &\dot{x}_2 = x_1 - x_1^3 - 0.1x_2 + u \label{eq:Duffing_oscillator}
\end{align}
\begin{figure}[ht]
  \centering
  \includegraphics[width=1\linewidth]{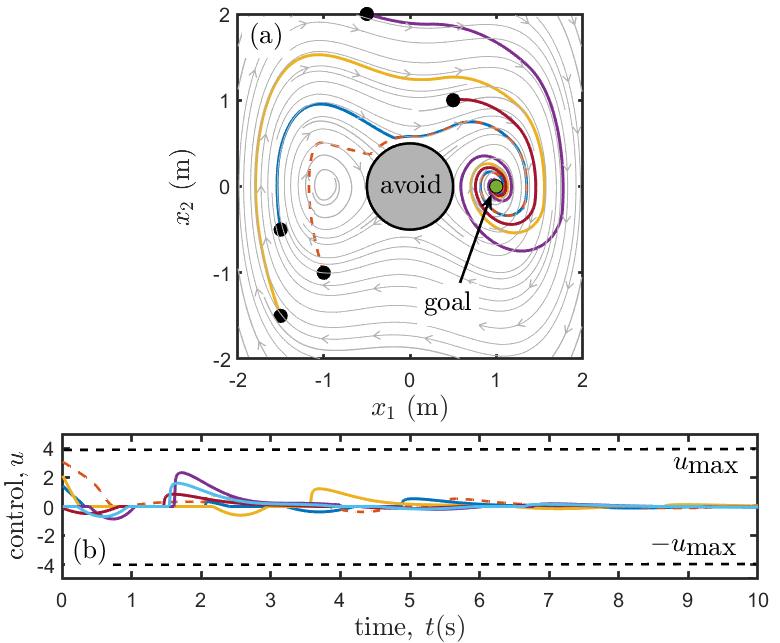}
\caption{(a) Trajectories converging to the target (green) while avoiding the unsafe set (gray), (b) Control plots for the controlled Duffing oscillator dynamics obtained by solving QP-CDF.}
\label{fig:Duffing}
\end{figure}

Fig.~\ref{fig:Duffing} provides a safe navigation control trajectory for the Duffing oscillator by solving QP-CDF. The density function is constructed based on equations \eqref{eq:bump} and \eqref{eq:rho_S}. The functions used to describe the obstacle set are given by $c(\bx):= \|\bx - o\|^2 - r_{1}^2$ where $o$ and $r_{1}$ are the center and radius of the circular obstacle respectively. Similarly, the function used to define sensing region is given by $b(\bx):= \|\bx - o\|^2 - r_{2}^2$ where $r_{2}$ is the radius of the sensing region of the circular obstacle. Here, the obstacle radius is 0.5 units and the sensing radius is 0.7 units with its center located at $[0,0]$. The matrix $P$ used to construct the function $V(\bx)$ is obtained by linearizing the dynamics around the target point and solving the algebraic Riccati equation with the identity matrix as the state and control gain matrix. Here, the control bounds are $\pm 2$. It can be observed that the control action obtained from the QP-CDF is minimally invasive. The system trajectories follow along the vector field of the Duffing oscillator when it is away from the obstacle. The control from the QP-CDF is mainly active near the obstacles by making the system trajectories drive away from the obstacles while trying to converge to the target set.

\subsection{Dubin car model}
Let us consider the Dubin car model as follows:
\begin{subequations}\label{eq:unicycle}
    \begin{align}
        \dot{x}_1 &= v\cos(\theta) \label{eq:unicycle_1}\\
        \dot{x}_2 &= v\sin(\theta) \label{eq:unicycle_2}\\
        \dot{\theta} &= \omega \label{eq:unicycle_3}
    \end{align}
\end{subequations}
where $u$ and $\omega$ are the control inputs. We assume that the obstacles to be avoided are only present in the $x_1 - x_2$ space. In this example, we first reformulate \eqref{eq:unicycle_1}-\eqref{eq:unicycle_2} in the form of single integrator dynamics as given in \eqref{eq:unicycle_si} and solve for $u_1$ and $u_2$ using the QP-CDF given in \eqref{eq:QP-CDF}. These control values will be then utilized in the design of $v$ and $\omega$. The single integrator dynamics in the $x_1 - x_2$ space is as follows:
\begin{subequations}\label{eq:unicycle_si}
    \begin{align}
        \dot{x}_1 &= u_1\label{eq:unicycle_si_1}\\
        \dot{x}_2 &= u_2 \label{eq:unicycle_si_2}
    \end{align}
\end{subequations}

The safe control for \eqref{eq:unicycle_si} is obtained by solving the QP-CDF given in \eqref{eq:QP-CDF}. Here, $c_k(\bx) := \|\bx - o_k\|^2 - r_{1k}^2$ and $b_k(\bx) := \|\bx - o_k\|^2 - r_{2k}^2$ where $o_k$, $r_{1k}$ and $r_{2k}$ are the center, radius and sensing radius of the $k^{th}$ circular obstacle. The control obtained as the solution of the QP-CDF can be used to calculate $u$ and $\Tilde{\theta}$ as follows:
\begin{align}
    v = \sqrt{u_1^2 + u_2^2},\;\;\;
    \Tilde{\theta} = \tan^{-1} \left(\frac{u_2}{u_1} \right)\label{eq:uni_theta_tilde}
\end{align}

\begin{figure}[ht]
  \centering
  \includegraphics[width=1\linewidth]{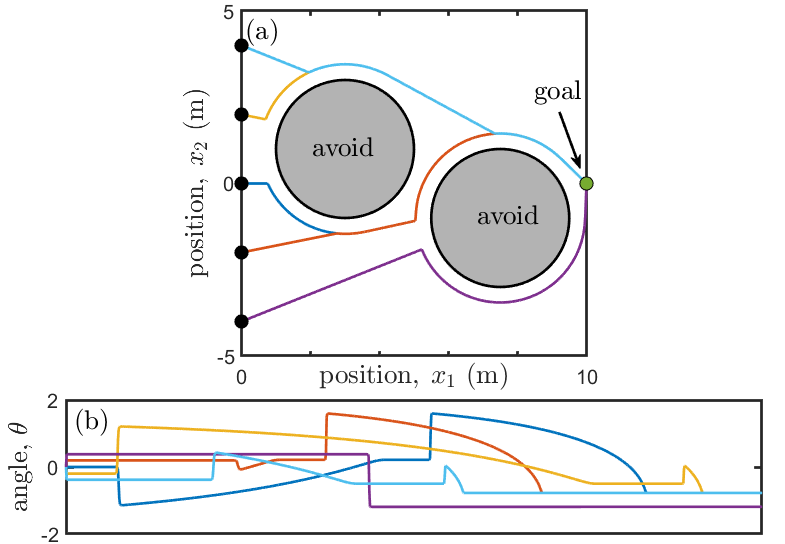}
\caption{(a) Trajectories in the $x_1 - x_2$ space converging to the target (green) while avoiding unsafe sets (gray), (b) state trajectory plot of $\theta$ for safe navigation.} 
\label{fig:unicycle}
\end{figure}

Now, the control $\omega$ needs to be designed such that $\theta - \Tilde{\theta}$ tends to zero.
Therefore, we consider the Lyapunov function given by $0.5(\theta-\Tilde{\theta})^2$ which gives us the following control law for $\omega$:
\begin{align}
    \omega = \dot{\Tilde{\theta}} - k(\theta-\Tilde{\theta}) \label{eq:uni_omega}
\end{align}
for some $k>0$. Fig.~\ref{fig:unicycle} provides a safe navigation trajectory for the unicycle model in $x_1 - x_2$ space. The obstacle radius is 2 units and the sensing radius is 2.5 units with the center for the two obstacles located at $[3,1]$ and $[7.5,-1]$ respectively. The value of gain $k$ is chosen to be 10. The control action mainly drives the system trajectories away from the obstacle set while converging on the target set.
\section{Conclusion}
The problem of safe navigation of nonlinear dynamical systems is considered. We introduce the notion of CDF as an extension to the density-based controllers. The CDF-based safety constraints are inspired by the occupancy-based interpretation of the measure associated with the density function. We then formulate the navigation problem with safety constraints as a QP using CDF. The existing approach for navigation problems includes CBF-based QP. The advantage of CDF-based QP over CBF-based QP is that both the convergence and safety constraints can be imposed using CDF. Finally, we provide simulation results for safe navigation of the Dubin car model and Duffing oscillator to showcase the validity of CDF-based QP.   
\section{Appendix}
The proof of Theorem \ref{thm:1} relies on the following Lemma.
\begin{lemma}\label{lemma_1}
    If
    \begin{align}
        \int_0^{\infty}\int_{\bX_0}\mathds{1}_{\mathcal{U}}(s_t(\bx)) d\bx \;dt=0 \label{eq:lemma_1}
    \end{align}
    then
    \begin{align}
        \int_{\bX_0}\mathds{1}_{\mathcal{U}}(s_t(\bx)) d\bx = 0 \;\;\;\forall t\ge0 \label{eq:lemma_2}
    \end{align}
    i.e., the amount of time system trajectories spend in set $\mathcal{U}$ starting from the positive measure set $\bX_0$ is equal to zero.
\end{lemma}
 \textbf{Proof:} 
The proof is done by the method of contradiction. Let us assume \eqref{eq:lemma_2} is not true. Then there exists a time $\bar t$ such that 
$$ \int_{\bX_0}\mathds{1}_{\mathcal{U}}(s_{\bar t}(\bx)) d\bx   >0.$$
Now, from the continuity of the solution of the differential equation, there exists a $\Delta$ such that
$$ \int_{\bar t}^{\bar t+\Delta}\int_{\bX_0}\mathds{1}_{\mathcal{U}}(\bs_t(\bx)) d\bx \;dt >0$$
Therefore,
\begin{align*}
    0 <&  \int_{\bar t}^{\bar t+\Delta}\int_{\bX_0}\mathds{1}_{\mathcal{U}}(\bs_t(\bx)) d\bx \;dt \\
    \leq& \int_0^{\infty}\int_{\bX_0}\mathds{1}_{\mathcal{U}}(s_t(\bx)) d\bx \;dt=0.
\end{align*}
Hence, we arrive at a contradiction.

\textbf{Proof of Theorem \ref{thm:1}:}
Let us consider the following:
\begin{align}
    \nabla \cdot (\bff (\bx)\rho + \bg(\bx)\bu\rho) = h(\bx) \label{eq:proof_1}
\end{align}
where $h(\bx) \ge 0$ and $h(\bx)\ge \lambda >0$ for $\bx \in \bX_0$. Now, through the method of characteristics, the function $\rho(\bx)$ can be written as follows:
\begin{align}
    \rho(\bx) = \int^{\infty}_0 h(s_{-t}(\bx))\left | \frac{\partial s_{-t}(\bx)}{\partial \bx}\right|dt,\label{eq:proof_2}
\end{align}
where $|\cdot|$ represents the determinant. This can be easily verified by simple substitution of \eqref{eq:proof_2} in \eqref{eq:proof_1} and using the fact that
\begin{align}
    \lim_{t\rightarrow \infty}h(s_{-t}(\bx))\left | \frac{\partial s_{-t}(\bx)}{\partial \bx}\right| = 0 .\label{eq:proof_3}
\end{align}
The limit in \eqref{eq:proof_3} is the consequence of $\rho(\bx)$ being bounded in $\Bar{D}$ and using Barbalat's Lemma. The term inside the integral in \eqref{eq:proof_2} can be written using the linear Perron-Frobenius (P-F) operator as follows:
\begin{align}
    [\mathbb{P}_t h](\bx) = h(s_{-t}(\bx))\left | \frac{\partial s_{-t}(\bx)}{\partial \bx}\right|.\label{eq:proof_4}
\end{align}
Therefore,
\begin{align}
    \rho(\bx) = \int^{\infty}_0  [\mathbb{P}_t h](\bx)dt .\label{eq:proof_5}
\end{align}
Now, utilizing \eqref{eq:proof_3}, we can write
\begin{align}
    \lim_{t \rightarrow \infty}[\mathbb{P}_t h](\bx) = 0 \implies \lim_{t \rightarrow \infty}[\mathbb{P}_t \mathds{1}_{\bX_0}](\bx) = 0 \label{eq:proof_6}
\end{align}
which follows because $h(\bx) \ge \lambda > 0\;\;\forall \bx\in\bX_0$ and using dominated convergence theorem. Here, $\mathds{1}_{\bX_0}$ represents the indicator function for $\bX_0$. Now, for any $A \subseteq \Bar{D}$, we have
\begin{align}
    \int_A [\mathbb{P}_t \mathds{1}_{\bX_0}](\bx)d\bx =& \int_{\Bar{D}}[\mathbb{P}_t \mathds{1}_{\bX_0}](\bx)\mathds{1}_{A}(\bx)d\bx\nonumber\\
    =&\int_{\Bar{D}}\mathds{1}_{\bX_0}(\bx)\mathds{1}_{A}(s_{t}(\bx))d\bx\label{eq:proof_7}
\end{align}
This can be observed by using the definition of P-F operator in \eqref{eq:proof_4} and doing the change of variables in integration such as $\by = s_{-t}(\bx)$ and $d\by=\left | \frac{\partial s_{-t}(\bx)}{\partial \bx}\right|d\bx$ and relabelling. The right hand side of \eqref{eq:proof_7} can be seen as follows:
\begin{align*}
    \int_A [\mathbb{P}_t \mathds{1}_{\bX_0}](\bx)d\bx = m\{\bx \in \bX_0 : s_t(\bx)\in A\}.
\end{align*}
Therefore, using \eqref{eq:proof_6}, we observe that
\begin{align*}
    0 =  \int_A [\lim_{t \rightarrow \infty}\mathbb{P}_t \mathds{1}_{\bX_0}](\bx)d\bx = m\{\bx \in \bX_0 : \lim_{t \rightarrow \infty}s_t(\bx)\in A\}.
\end{align*}
The above statement can be generalized for any measurable Lebesgue set $A \subseteq \Bar{D}$. Therefore,
\begin{align*}
    m\{\bx \in \bX_0 : \lim_{t \rightarrow \infty}s_t(\bx)\neq 0\} = 0.
\end{align*}
Now, from the construction of the density function, we know that $\rho(\bx)=0\;\forall\;\bx \in \mathcal{U}$. Therefore,
\begin{align}
    \int_{\mathcal{U}}\int^{\infty}_0[\mathbb{P}_t \mathds{1}_{\bX_0}](\bx)dt d\bx \le \int_{\mathcal{U}}\rho(\bx)d\bx = 0. \label{eq:proof_8}
\end{align}
Utilizing the Markov property of the P-F operator and the fact that indicator functions are non-negative functions, we can rewrite \eqref{eq:proof_8} as follows:
\begin{align}    \int_{\mathcal{U}}\int^{\infty}_0[\mathbb{P}_t \mathds{1}_{\bX_0}](\bx)dt d\bx  = 0. \label{eq:proof_9}
\end{align}
Now, doing the change of variables in integration such as $\by = s_{-t}(\bx)$ and $d\by=\left | \frac{\partial s_{-t}(\bx)}{\partial \bx}\right|d\bx$ and relabelling, the left-hand side of the \eqref{eq:proof_9} can be written as follows:
\begin{align}
\int_{\mathcal{U}}\int^{\infty}_0[\mathbb{P}_t \mathds{1}_{\bX_0}]&(\bx)dt d\bx = \int_{\Bar{D}}\int^{\infty}_0 [\mathbb{P}_t \mathds{1}_{\bX_0}](\bx)\mathds{1}_{\mathcal{U}}(\bx)dt d\bx \nonumber\\
=& \int_0^{\infty}\int_{\Bar{D}}\mathds{1}_{\bX_0}(\bx)\mathds{1}_{\mathcal{U}}(s_t(\bx)) d\bx \;dt\nonumber\\
=& \int_0^{\infty}\int_{\bX_0}\mathds{1}_{\mathcal{U}}(s_t(\bx)) d\bx \;dt=0.\label{eq:proof_10}
\end{align}
Now, from \eqref{eq:proof_10} and using Lemma \ref{lemma_1}, we can conclude the following:
\begin{align*}
    \int_{\bX_0}\mathds{1}_{\mathcal{U}}(s_t(\bx)) d\bx = 0 \;\;\;\forall t\ge0   
\end{align*}

\bibliographystyle{IEEETRAN}
\bibliography{reference,Umesh_ref}
\end{document}